\documentclass[11pt]{article}
\usepackage[]{amsmath,amssymb}
\usepackage{graphics,epsfig}

\begin{document}
\date{}
\title{Entropy localization and extensivity in the semiclassical black hole evaporation }
\author{H. Casini\footnote{e-mail: casini@cab.cnea.gov.ar} \\
{\sl Centro At\'omico Bariloche,
8400-S.C. de Bariloche, R\'{\i}o Negro, Argentina}}
\maketitle

\begin{abstract}
We aim to quantify the distribution of information in the Hawking radiation and inside the black hole in the semiclassical evaporation process. The structure of relativistic quantum field theory does not allow defining a localized entropy unambiguously, but rather forces to consider the shared information (mutual information) between two different regions of space-time. Using this tool, we first show that the entropy of a thermal gas at the Unruh temperature underestimates the actual amount of (shared) information present in a region of the Rindler space.  
Then, we analyze the mutual information between the black hole and the late time radiation region. A well known property of the entropy implies that this is monotonically increasing with time. This means that in the semiclassical picture it is not possible to recover the eventual purity of the initial state in the final Hawking radiation through subtle correlations established during the whole evaporation period, no matter the interactions present in the theory.  
We find extensivity of the entropy as a consequence of a reduction to a two dimensional conformal problem in a simple approximation. However, the extensivity of information in the radiation region in a full four dimensional calculation seems not to be guaranteed on general grounds. 
 We also analyze the localization of shared information inside the black hole finding that a large amount of it is contained in a small approximately flat region of space-time near the point where the horizon begins. This gives place to large violations of the entropy bounds.  We show that this problem is not eased by backscattering effects and argue that a breaking of conformal invariance is necessary to delocalize the entropy. 
  Finally, we indicate that the mutual information could lead to a way to understand the Bekenstein-Hawking black hole entropy which does not require a drastic reduction in degrees of freedom in order to regulate the entanglement entropy. On the contrary, a large number of field degrees of freedom at high energies giving place to a Hagedorn transition implements a natural distance cutoff in the mutual information, which may in consequence turn out to be bounded.
\end{abstract}

\section{Introduction} 
Since the early days of quantum mechanics it was recognized that the relativistic analog of the non-relativistic quantum theory of a single particle contains inconsistencies. As it turned out, the root of the problem is in the concept of a localizable particle, which ceases to be meaningful in the relativistic case. In this later, the localization process inevitably leads to the creation of particle-antiparticle pairs, and requires the simultaneous description of an infinite number of degrees of freedom. This is achieved by the quantum field theory (QFT), which is the natural synthesis of quantum mechanics and special relativity allowing for a sharp space-time localization of events. 

This proliferation of degrees of freedom due to Lorentz invariance and locality has many important consequences.   One of those is the presence of ambiguities when one attempts to define the entropy enclosed in a given volume. In order to define this entropy, consider a global state (density matrix) $\rho$. We can reduce it to the spatial set $V$ by tracing out the degrees of freedom lying outside $V$ 
\begin{equation}
\rho_V=\textrm{tr}_{-V} \rho\,.
\end{equation}
The density matrix $\rho_V$ is the localized state relevant for computing expectation values of the observables acting in $V$. However, its von Newman entropy (which in the case of the vacuum state $\rho=\left|0 \right\rangle\left\langle 0\right|$ is called entanglement or geometric entropy)
\begin{equation}
 S(V)=-\textrm{tr}\rho_V\log \rho_V \,,
\end{equation}
 is divergent \cite{tho,bom}. This is due to the infinite amount of entanglement present in the vacuum fluctuations around the boundary of $V$. The trace leaves one member of a virtual pair as a "real particle" inside $V$ while the other outside $V$ is traced over.   Due to their ultraviolet nature, 
 the divergent terms appearing in the entanglement entropy of the vacuum state also show up in the entropy of any finite energy state reduced to the region $V$.

In the text-book calculation of the entropy of a gas this problem is bypassed by imposing artificial boundary conditions on the region considered (i.e. a box with perfectly reflecting walls), and taking a state and a theory defined inside, avoiding in this manner the entanglement with the exterior region. Then, intensive quantities such as the entropy per unit volume are computed in the large volume limit. In this way one obtains results which are generally independent of the particular boundary conditions, since they contribute at most with a term increasing like the boundary area. Alternatively, one can get the same result for the mean entropy by imposing a cutoff to regularize the entropy of $\rho_V$ and then letting the volume go to infinity before removing the cutoff. A theorem   based on the strong subadditive property of the entropy guarantee the existence of this limit in translational invariant situations \cite{rr}. Therefore, in the context of special relativity, the localized entropy is tied to entropy density and to entropy extensivity.

However, the ambiguities are unavoidable when the large volume limit cannot be taken. This is the case if one is interested in the entropy radiated by a black hole, or the one inside a cosmological horizon.  Moreover, the existence of these types of horizons is inevitable when taking a large volume limit of homogeneous matter in the presence of the gravitational interaction, disregarding how weak this later is. This means that the black hole entropy problem is deeply related with the issue of state localization in relativity. 

Therefore, strictly speaking, in semiclassical gravity there is no well defined notion of localized entropy, not even in the sense which is allowed in special relativity.  However, it can be argued that in some situations the high energy modes responsible for the entropy divergences decouple from the low energy ones. This may manifest itself in the presence of different terms (area and volume growing terms for example) which can be isolated and approximately separated. Thus, if this decoupling is efficient, one can achieve the large volume limit before gravity becomes relevant. Still, the decoupling is not evident at least for diluted gases where the contribution of the boundary to the low energy modes is of the order of the entropy one is interested in evaluating. This is certainly the case of the Hawking radiation, and may even be relevant to the radiation escaping from a star in the far away region. 

In relation with the scenarios of horizon formation, several entropy bounds have been proposed in the literature \cite{eb,fmw}. These would forbid the entropy to grow faster than a constant times the boundary area of the region considered, giving a sharp expression to the problem of the entropy ambiguity. If a form of entropy bound holds in nature we certainly need to find new principles which guarantee the extensivity of the entropy. This may be part of the difficult task usually attributed to the holographic principle, which would be operative in quantum gravity \cite{holo}.

However,  we think this problem seems to require new ideas at low energies. Indeed, imagine we have come out with an expression for the entropy of a set $V$ in flat empty Minkowski space. One may wonder what this entropy would actually do: describe the empty space seen by an inertial observer or rather the entropic radiation measured by an accelerated one.  This phenomenon of observer dependence of the entropy has been pointed out in \cite{maro}. This suggests that the entropy ambiguity is somehow required by relativity and that it cannot be simply eliminated.  In other words, it should not be possible to solve the localization problem just by producing a unique prescription to compute the entropy corresponding to a region in the classical limit of the space-time (for a different opinion see for example \cite{wwaa}). This would be a new  manifestation of the well known fact that a proper definition of entropy requires the specification of a coarse graining. 

In this paper we explore the question of entropy localization and extensivity at the semiclassical level. The entropy is then necessarily ambiguous. Instead of it we search for quantities which may replace the notion of localized entropy in QFT. We find that there is a natural and almost unique answer, which is given by the mutual information between two different non-intersecting regions. It has the meaning of the information shared by these two regions and it is regularization independent and well defined mathematically. In some sense it can be considered entropy in which a specific form of coarse graining has been imposed by the structure of the theory. We propose it as the correct tool to evaluate the entropy carried by the vacuum fluctuations.

Two properties of the mutual information are relevant in this context: it is monotonically increasing with the size, and it gives a lower bound to the entropy on the regions involved. This later property was used in an attempt to properly define and test the entropy bounds at the semiclassical level \cite{hor}.  
Here, we study the mutual information in the Schwarzschild black hole evaporation problem. In a simple two dimensional approximation we find, somewhat surprisingly, extensivity for the mutual information on the radiation region.  For the full four dimensional case the calculation is more involved, and we could not obtain a closed result. We point out however that the localization properties of the entropy in the Hawking radiation are far from clear, and in principle it can be distributed very differently than the radiated energy.  

We also study the localization of the entropy inside the black hole, and find that it is strongly concentrated near the point where the horizon begins. We find that the amount of shared entropy there exceed the Bekenstein-Hawking entropy (or black hole entropy) $S_{BH}={\cal A}/(4G)$. We discuss this result, and use the mutual information in trying to give a semiclassical statistical interpretation of the black hole entropy and the generalized second law.

\section{Mutual information}

The localized entropy has divergences which cannot be renormalized in the context of QFT. However, these have a nice geometrical structure which leads to a very specific way to subtract them away.

For the sake of analyzing the ultraviolet divergent terms we can use a pure global state. Thus we have the symmetry property  $S(A)=S(-A)$, where $-A$ is the set complementary to $A$ on a global Cauchy surface. One can freely change the cutoff and make it point dependent (for example using a non homogeneous lattice), but this equality has to be maintained. This implies that the divergent terms, which being ultraviolet are local, must depend only on the boundary surface of $A$, which is shared with $-A$. They are also extensive on the boundary because the entropy adds for independent variables. On a more technical level, the entanglement entropy $S(A)$ corresponding to the spatial set $A$ is proportional to the variation of the Euclidean free energy with respect to the introduction of  conical singularities at the boundary  \cite{holz}. More precisely 
\begin{equation}
S(A)=\lim_{\alpha\rightarrow 1} \frac{\log Z(A,\alpha)- \alpha \log Z(A,1)}{1-\alpha } ,\label{alpha}
\end{equation}
where $Z(A,\alpha)$ is the Euclidean partition function of the theory on a space with conical singularity of angle $2 \pi \alpha$  at the boundary of $A$ (for $\alpha=1$ there is no singularity). 
As a result, the ultraviolet divergent terms in the entropy $S(A)$ are local on the set boundary, where the manifold is singular.  In $d$ spatial dimensions, we then have an expansion of the form \cite{ch2} (see also \cite{conical})
 \begin{equation}
 S(A)=g_{d-1}[\partial A]\,\epsilon^{-(d-1)}+...+ g_1[\partial A]\,\epsilon^{-1} + g_0[\partial A]\,\log (\epsilon)+ S_0(A)\,,   \label{div}
 \end{equation}
 where $S_0(A)$ is a finite part, $\epsilon$ is a short distance cutoff, and the $g_i$ are local and extensive  functions on the boundary $\partial A$, which are homogeneous of degree $i$. The leading divergent term coefficient  $g_{d-1}[\partial A]$ is  proportional to the $(d-1)$ power of the size of $A$. This fact was recognized since the earliest papers on the subject and it was thought as an area law for the entanglement entropy \cite{bom,dos}, which mimic and pointed to a common origin with the black hole entropy formula. However, $g_i$ for $i> 0$ depends on the regularization procedure and $g_{d-1}$ is not proportional to the area if this later is not rotational invariant. This term is not physical within QFT since it is not related to continuum quantities (this divergence might be absorbed in the renormalization of $G$ in quantum gravity \cite{tres}). The same can be said of the subleading divergent terms, excepting the logarithmic one, which has universal (cutoff independent) coefficient \cite{ch2}.  

Thus, to produce a finite physical quantity out of $S(A)$ we have to subtract the boundary contributions. As a bonus, this would automatically eliminate the problematic negative contact terms in the entropy which appear for theories containing gauge fields or non-minimally coupled scalars \cite{contact}. Two different ways to make this subtraction naturally suggest themselves. The first one consists in subtracting the entropies for two different  global states, $S(\rho_1^A)- S( \rho_2^A)$ as done in \cite{maro}. The ultraviolet behavior should not depend on the state and must cancel in this combination. A variation of this is  
\begin{equation}
S(\rho_1^A | \rho_2^A)=\textrm{Tr} \rho_2^A \log \rho_2^A - \textrm{Tr} \rho_2^A \log \rho_1^A\,.
\end{equation}
This particular combination is a well known quantity called the relative entropy $S(\rho_1 | \rho_2)$ between two states $\rho_1$ and $\rho_2$. It has good properties which resemble the ones for the entropy. In particular $S(\rho_1^A | \rho_2^A)\ge 0$. It can be argued that this inequality for the particular case when $\rho_1^A$ is the reduced density matrix of the vacuum provides a version of the Bekenstein bound in flat space which is well defined and true. 

There are two different kinds of problems with this type of regularization. The first one occurs even for a fixed non dynamical metric. One uses here a reference state to define the local entropy of other state. This may be natural in flat space where one may subtract the entropy of the vacuum, but in the general case where there is no standard vacuum nothing dictates the election of a particular reference state.  
The second problem is that when the metric is dynamical two different states live in different geometries. This introduces an ambiguity in the region which has to be considered for each density matrix. To see how bad this uncertainty can be, consider a nearly flat case with an object of mass $M$ enclosed in a region $V$ of typical size $R_V\gg MG$, and compare its entropy to the entanglement entropy of the vacuum in $V$. The uncertainty in the area produced by the Newtonian potential is $\delta {\cal A }\sim \frac{MG}{R_V} {\cal A}$. The corresponding entropy uncertainty considering a Planck scale cutoff is $\delta S \sim \delta {\cal A} G^{-1}\sim  M R_V$, which has to be much smaller than the actual renormalized value of the entropy $S$ obtained from the subtraction of the entanglement entropy of the vacuum. This gives $S\gg M R_V$, which is just the opposite of the Bekenstein bound~\footnote{This is somehow a reverse argument to the one given by Bousso in \cite{bou}, which  proves the  Bekenstein bound from the FMW bound \cite{fmw}.}. 
This problem is eased if we take a cutoff distance larger than Planck scale, but still this type of subtractions do not lead to well defined localized entropy in the semiclassical case.  

Thus, we need a function of one state rather than two. To eliminate the divergences using only one global state we crucially take into account that they are extensive on the boundary. We then consider two different nonintersecting sets and compute \cite{ch0}
\begin{equation}
I(A,B)=S(A)+S(B)-S(A\cup B)=S( \rho_A\otimes \rho_B | \rho_{A\cup B})\,.\label{fff}
\end{equation}
We pay the price that our entropy function depends now on two entries but this is a one state function and a regularization independent measure of entanglement. Note that the divergent boundary terms get subtracted in $I(A,B)$. 
 In consequence, finite terms in the function $S(A)$ which are local on the boundary, like the area term, should be considered like gauges, since they do not modify the physical quantities like $I(A,B)$.    
To define $S(A\cup B)$ the two spatial sets $A$ and $B$ have to be spatially separated to each other in order that $A\cup B$ is included in a Cauchy surface.    
In flat space, if the global state is the vacuum, $I(A,B)$ is a characteristic property of the particular QFT theory. 

The locality of QFT leads naturally to the formula (\ref{fff}) in order to extract the universal part contained in the entropy. Curiously, the same quantity has independently been studied long before in statistics and information theory. It was named mutual information and has the meaning of a measure of the information shared between the two different systems (see for example \cite{dieci}).

For a pure global state we have $S(A)=S(-A)$, and it follows from (\ref{fff}) that $I(A,-A)=2S(A)$. Thus, in QFT when $B$ tends to cover the space outside of $A$, the mutual information $I(A,B)$ tends to infinity, somehow reproducing the divergences which are present in $S(A)$, but in  a cutoff independent way. In fact, in general divergences appear in $I(A,B)$ whenever the boundaries of the two sets touch each other.

On the mathematical side, the mutual information can be defined rigorously in QFT when $A$ and $B$ are separated by a non zero distance (see however section 5 below). In the axiomatic approach there is no local Hilbert space of states attached to each region $O$, but rather a local algebra ${\cal A}_O$ of operators generated by the quantum fields. Then the mutual information can be defined using the Araki formula \cite{ara} for the relative entropy of a pair of states in an operator algebra (here $\rho_{A\cup B}$ and $\rho_A \otimes \rho_B$ in the algebra ${\cal A}_{A\cup B}$ \cite{aqft}).  

Thus, when we write $I(A,B)$ the sets $A$ and $B$ just identify the corresponding local operator algebras in the QFT. This gives a bonus. Due to causality, the algebras in QFT are generally attached to causally complete regions. Typical examples are diamond shaped regions as shown in Figure 1. This means that 
\begin{equation}
I(A,B)=I(A^\prime ,B^\prime )\equiv I(\hat A, \hat B)\,,
\end{equation}
 where  $A^\prime$ and $B^\prime$ are any spatial sets which are Cauchy surfaces for the same D-dimensional set as $A$ and $B$. In other terms, the mutual information is a function of the equivalence classes of spatial sets having the same causal completion, or what is the same, a function of causally complete regions. This forces us to think in space-time terms. The same requirement for the entropy would be that $S(A)=S(A^\prime)$
 for equivalent spatial sets $A$ and $A^\prime$. This can be guessed true from the unitarity of evolution, but the application of this relation may be hindered by the necessity of regularization. 

\begin{figure}
\centering
\leavevmode
\epsfysize=4cm
\epsfbox{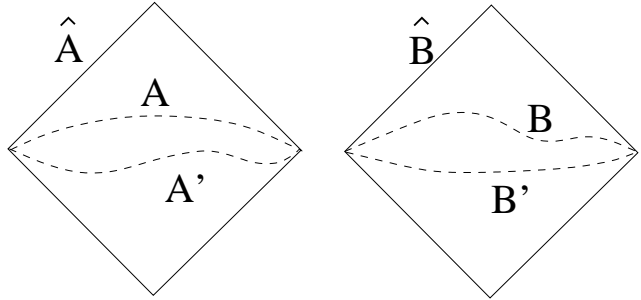}
\bigskip
\caption{The spatial surfaces $A$ and $A^\prime$ correspond to the same causally complete region $\hat{A}$ (diamond shaped set), while $B$ or $B^\prime$ are Cauchy surfaces for the causally complete region $\hat{B}$.}
\label{f1}
\end{figure}

As an application of the mutual information, we may give an argument to the effect that the entanglement entropy cannot be finite in QFT, independently of the details of the high energy sector. In \cite{cas} it was shown that if the entanglement entropy is finite and Lorentz invariant it must be exactly proportional to the boundary area. The proof is based on the strong subadditive property of the entropy, and involves crucially short and long distances of arbitrary size in Minkowski space. An entropy proportional to the area means that the mutual information is identically zero, $I(A,B)\equiv 0$. This simply cannot hold since the mutual information is a low energy quantity which is non zero in QFT (see the examples below). The resolution of this tension in QFT implies divergent entanglement entropy.  

Next we review some mathematical properties of the mutual information which are appropriate to an information measure (for reviews of the properties of entropy and related quantities see for example \cite{rev}). We also show some explicit examples in flat space-time. In the following sections we begin to investigate to what extent it can be considered as a useful quantity which replaces the localized entropy for relativistic theories.

\subsection{General mathematical properties of I(A,B)}
In a general quantum system (not necessarily the local reduced density matrices in QFT), the function $I(A,B)$ is symmetric, dimensionless and positive 
\begin{equation}
I(A,B)\ge 0 \,.
\end{equation}
 For product state ($A$ and $B$ statistically independent), $\rho_{A\cup B}=\rho_A\otimes \rho_B$, the mutual information  is zero $I(A,B)=0$. 
It has the remarkable property of monotonicity with respect to size,
\begin{equation}
 I(A,B)\le I(A,C), \hspace{2.5cm} B\subseteq C \,.\label{ine1}
\end{equation} 
This means that the mutual information is smoothly varying with the set, and that the sharpness in the definition of  the boundaries is not very relevant to it. This property is a consequence of the strong subadditivity of the entropy \cite{ssa}
\begin{equation}
S(A)+S(B)\ge S(A\cup B) +S(A\cap B)\,,\label{ssaa}
\end{equation}
but hold in the more general case in which $S(A)$ cannot be defined, as in QFT \cite{ara}.  
We have also the general inequality   
\begin{equation}
I(A,B)\le 2 \min(S(A),S(B))\,.\label{fofo}
\end{equation}  

We can define a form of tripartite information (somehow information shared between $B$ and $C$ with respect to $A$)
\begin{equation}
I(A\vert B,C)=I(A,B)+I(A,C)-I(A,B\cup C)\,.\label{non}
\end{equation}
Because of the definition (\ref{fff}) this function has complete permutation symmetry
\begin{equation}
I(A,B,C)\equiv I(A\vert B,C)=S(A)+S(B)+S(C)-S(A\cup B)-S(A\cup C)-S(B\cup C)+S(A\cup B \cup C)\,.
\end{equation}
From (\ref{non}) we see that $I(A,B,C)$ measures the degree of non-extensivity of the mutual information, in the same way as $I(A,B)$ measures the degree of non-extensivity of $S$. However, $I(A,B,C)$ can have both signs. It is zero  when the total state is pure but the general condition which makes the mutual information extensive is unknown to the author. 

\subsection{Some examples in Minkowski space}
The explicit calculation of the entanglement entropy and the mutual information involves a partition function in a non-trivial manifold which is generally  difficult to evaluate. However, in some particular cases exact results can be obtained. Here we recall some known results with interest for the following sections.

From the point of view of QFT the mutual information can be thought as a kind of correlator of set variables. One expects that for separating distances between $A$ and $B$ which are large with respect to their sizes $I(A,B)$ decays exponentially fast for massive theories and as a power of distance for massless ones. For free theories it can be shown explicitly using the results of \cite{peschel} and a lattice regularization that the mutual information decays as the square of the field correlator in the large distance limit and for the space-time dimension $D> 2$ \cite{veinticinco}. For massless fermions that means 
\begin{equation}
I(A,B)\sim \frac{G_F(A,B)}{R^{2(D-1)}}\label{sus1}\,,
\end{equation} 
while for free massless scalars it is  
\begin{equation}
I(A,B)\sim \frac{G_B(A,B)}{R^{2(D-2)}}\label{sus2}\,.
\end{equation} 
The functions $G(A,B)$ have the dimensions which make $I(A,B)$ dimensionless, and are monotonously increasing with $A$ and $B$. They depend on their relative orientation but are independent of their mutual distance $R$. For massive fields the power of $R$ is replaced by an exponential. 

The opposite situation is one with two sets with boundaries very near to each other. The leading term can be inferred from dimensional analysis in this case. Take two sets $A$ and $B$ with parallel faces of area ${\cal A}$, at a distance $L$ much smaller than the other dimensions of the geometric arrangement. We  then have \cite{ch4}
\begin{equation}
I(A,B)\simeq \kappa\, \frac{{\cal A}}{L^{D-2}}\,.\label{tuy}
\end{equation}
 The dimensionless function $\kappa$
for free fields has been computed explicitly. For massive fields it is 
\begin{equation}
\kappa(t) =   \frac{ t^{D-2}}{2^{D-3}\,\pi^{\frac{D-2}{2}}\,\Gamma\left(\frac{D-2}{2}\right)}\,\,\int_t^\infty dy_1\, y_1^{-(D-1)} \int_0^\infty d y_2\, \, y_2^{D-3}\, c\left(\sqrt{y_1^2+y_2^2}\right)\,,\label{masapan}
\end{equation}
where $t=m L$, and $m$ is the fields mass. In the massless limit it reduces to a constant  
\begin{equation}
\kappa =  \left( (D-2)\,2^{D-3}\,\pi^{\frac{D-2}{2}}\,\Gamma\left(\frac{D-2}{2}\right)\right)^{-1} \int_0^\infty dy \,\,y^{D-3}\, c(y) \,.\label{ttt}
\end{equation}
 The function $c(t)=(n_B c_B(t)+n_F c_F(t))$, where $n_B$ and $n_F$ are the multiplicity of the bosonic and fermionic degrees of freedom, and $c_B(t)$ and $c_F(t)$ are the two dimensional entropic functions corresponding to boson and fermion fields \cite{ch4,ch3}. These are defined by the formula $c(mL)=L\,dS(L)/dL$, where $S(L)$ is the entanglement entropy corresponding to an interval of length $L$ in $1+1$ dimensions \cite{ch0}. The subindex $B$ corresponds to a two dimensional real scalar field and $F$ to a Majorana fermion one. These functions are given in terms of solutions of Painleve V nonlinear ordinary differential equations \cite{ch5}.

The analytic expression for the entropy has also been obtained for conformal field theories in two dimensions, where the $n$-sheeted space can be conveniently transformed. In this case the entropy for a set formed by $p$ disjoint spatial intervals is \cite{calcar}
\begin{equation}
S=\frac{C_V}{3}\left( \sum_{i,j} \log d\left( a_i,b_j \right)-\sum_{i<j}
\log d\left( a_i ,a_j \right)-\sum_{i<j}\log d\left( b_i ,b_j \right)-p\,\log\epsilon \right)\,,\label{confor}
\end{equation}
where $a_i$ and $b_i$ are the left and right endpoints of the $i^{\textrm{th}}$ interval, $\epsilon$ is the ultraviolet cutoff, $C_V$ is the Virasoro central charge, and $d$ is the distance function in Minkowski space. 
Remarkably, this specific form of the entropy gives mutual information which is extensive \cite{ch3}, 
\begin{equation}
I(A,B\cup C)=I(A,B)+I(A,C)\,,\label{ex}
\end{equation}
or we could also say distributive, $I(\cup_i A_i ,\cup_j B_j)=\sum_{i,j} I(A_i,B_j)$. 
 This property does not hold in general for massive theories or massless ones in higher dimensions as shown by  (\ref{sus2}). 
For two intervals $A$ and $B$ on the same plane of sizes $l_A$ and $l_B$, separated by a distance $l_C$, we have 
\begin{equation}
 I(A,B)=\frac{C_V}{3} \log \left( \frac{(l_A+l_C) (l_C+l_B) }{l_C (l_A+l_B+l_C)} \right)= \frac{C_V}{3} \log \left(\eta \right)\label{jjj}\,,
\end{equation}
which is explicitly cutoff independent and conformal invariant since it is expressed in terms of a cross ratio $\eta$  \cite{ch0}. The entropy and the mutual information in the conformal case can be written in null coordinates, $u=x-t$ and $v=t+x$, as a sum of two independent terms, one for each coordinate, representing the independent contributions from the right and left moving chiralities. For example, for two intervals we have (not necessarily on the same spatial plane) 
\begin{equation}
 I(A,B)=\frac{C_V}{6} \log \left( \frac{(u_A+u_C) (u_C+u_B) }{u_C (u_A+u_B+u_C)} \right)+  \frac{C_V}{6} \log \left( \frac{(v_A+v_C) (v_C+v_B) }{v_C (v_A+v_B+v_C)} \right)\label{chira}\,,
\end{equation}
where $u_X$ and $v_X$ are the null coordinate sizes of $X$, $l_X=\sqrt{u_X\,v_X}$.  While (\ref{confor}) shows that the functional form of the entanglement entropy is the same (excepting a global constant) for all two dimensional conformal field theories, this is not the case for conformal theories in more dimensions (see (\ref{sus1}) and (\ref{sus2})).
  
\section{Rindler space}
 As an introduction to mutual information in black hole backgrounds let us first focus on the simpler case of a Rindler space, where we can use the results of the previous Section. 
 The entanglement entropy of the Rindler wedge is of course divergent, proportional to the logarithm of the cutoff in two dimensions, and to ${\cal A}/\epsilon^{D-2}$ in $D$ space-time dimensions, where ${\cal A}$ is the transversal area of the wedge. 

 Then consider the mutual information between the Rindler wedge $A$ and another set $B$ exterior to it, as shown in figure 2. 
 Let us take a conformal theory since in the case there is mass $I(A,B)$ will rapidly go to zero with the mutual  distance $l_C$. 
 
 First take the space-time dimension $D=2$ and $B$ a segment of length $l_B$. 
 The equation (\ref{jjj}) for $l_A\rightarrow \infty$ gives 
\begin{equation}
I(A,B)=\frac{C_V}{3} \Delta \nu \,,\label{veintiuno}
\end{equation}
where $\nu =\text{arcth} (t/x)$ is a boost parameter, and 
 \begin{equation}
\Delta \nu=\log\left(\frac{l_B+l_C}{l_C}\right)\,
\end{equation}
 is the $\nu$ interval corresponding to the set $B$. We can also write 
 \begin{equation}
 \Delta \nu=a \Delta \tau\,,
 \end{equation}
  where $\tau$ is
the proper time measured
by an accelerated
observer with constant acceleration $a$.
Thus, according to (\ref{veintiuno}) the mutual information is proportional to the diamond
Rindler's time $\Delta \tau$ times a constant "flux of information" $\frac{C_V}{3} a$ due to a constant flux of Unruh radiation  \cite{ch0}. This gives an interesting meaning  to the surprising extensivity property (\ref{ex}) for the mutual information of conformal fields.  

Let us explore  in more detail this relation between mutual information and Unruh radiation. The Unruh temperature
 measured by an accelerated observer of constant acceleration $a$ is $T=a/(2\pi)$ and this observer is at a constant distance $1/a$ from the wedge vertex. Thus, in general we have a temperature $T=1/(2\pi x)$, varying with space-time point, where $x$ is the distance to the wedge. In fact, the density matrix for the half space is thermal with respect to the boost operator $K$  
 \begin{equation}
 \rho\sim e^{-2 \pi K}\label{seee}
 \end{equation} 
 rather than the Hamiltonian, with a dimensionless temperature $(2\pi)^{-1}$. 
On the other hand, in the large volume limit the entropy per unit volume for a conformal theory is \cite{car}
\begin{equation}
\frac{dS}{dx}=\frac{\pi C_V}{3} T.
\end{equation}
 If we use (heuristically) this expression for computing the entropy of the radiation in $B$ we have
\begin{equation}
\int_{l_C}^{l_C+l_B} dS=\int_{l_C}^{l_C+l_B} \frac{\pi C_V}{3} \frac{dx}{2\pi x}=\frac{C_V}{6}\label{heu} \log\left(\frac{l_C+l_B}{l_C}\right)\,,
\end{equation}  
which is exactly one half of the mutual information. 

The factor two could be ascribed to the fact that when the global state $\rho_{AB}$ is pure (that is when $A\cup B$ tends to the whole space) the entropy appears duplicated in the mutual information, $I(A,B)=S(A)+S(B)=2\, S(A)$. 
 However, the coincidence of the mutual information and twice the entropy heuristically contained in the thermal Unruh radiation is a peculiarity of the two dimensional conformal case, where the wedge can be transformed to a Euclidean cylinder \cite{holz}. In general there is no justification for using the large volume limit entropy density in the calculation above.

\begin{figure}[t]
\centering
\leavevmode
\epsfysize=5.5cm
\epsfbox{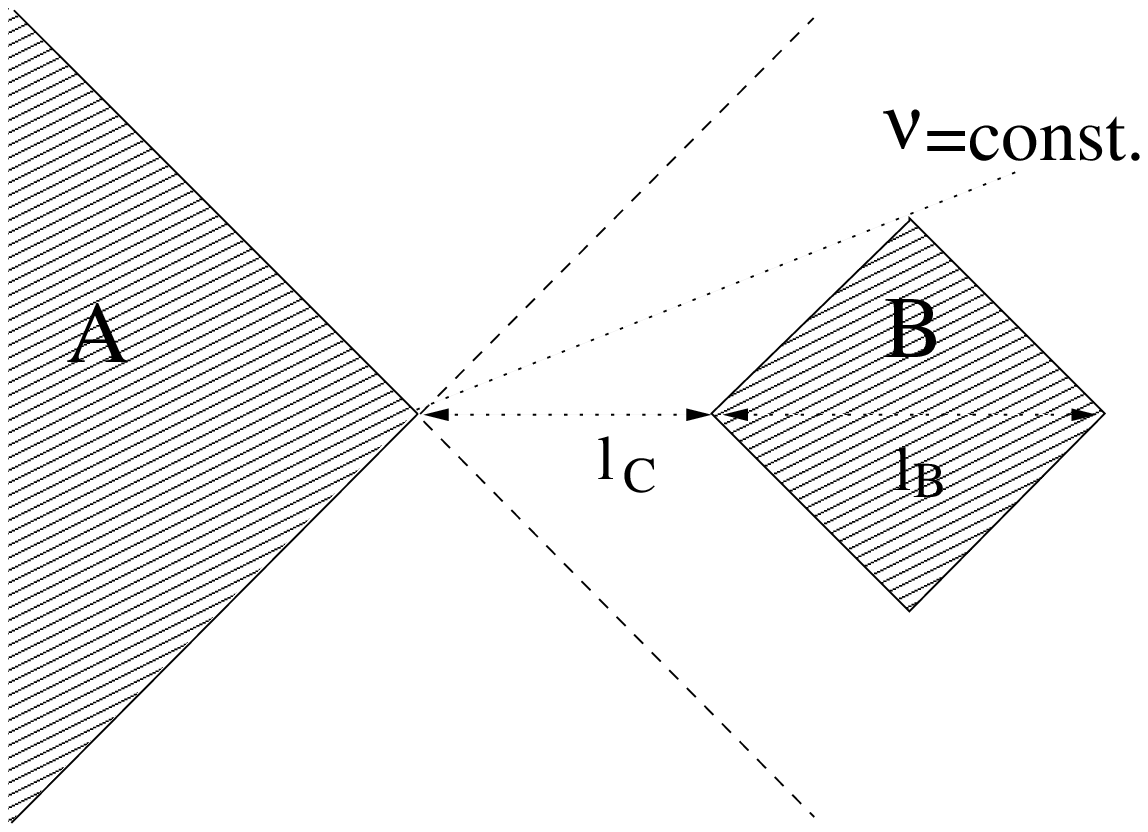}
\bigskip
\caption{The wedge $A$ and diamond set $B$ in the exterior region. The curves of constant $\nu$ are
 lines through the origin.}
\end{figure}

On the other hand, for a theory in two dimensions, the proportionality of $I(A,B)$ with Rindler's time can be  understood in the large $\Delta \nu$ limit.
 In fact, in this limit we can even define an entropy density without resorting to the mutual information. This is because the contribution of the boundary of $B$ to the entropy is constant, since it always consists of two points, while the Rindler time during which the radiation is collected grows indefinitely. 
Thus, the boundary contribution can be neglected for the mean entropy per unit $\nu$. More formally, the extensivity in this limit can be shown as a consequence of the boost symmetry. In order to see this we follow the argument which was used in \cite{rr} for probing the existence of  a mean entropy in the large volume limit. 
There the authors considered a translational invariant one dimensional system, with   $S(l)$ the entropy of a length $l$ interval. Using the SSA inequality (\ref{ssaa}) between one segment and its translation by an infinitesimal amount one gets that $S(l)$ is a (positive) concave function of $l$, $S^{\prime\prime}(l)\le 0$. This automatically guarantee that $S(l)/l$ approaches a constant for large $l$. In the present case the same argument can be used for $S(B)$ (or $S(A\cup B)$) as a function of the $B$ Rindler's time $\Delta \nu$, where now the role of translations is played by the boost symmetry. 
  Note we have to use the SSA inequality here between sets which are not on the same Cauchy surface but rather they are  time-like displaced by a boost \cite{cas}. This is not an impediment for applying the SSA relation for one dimensional conformal theories since one can interchange the meaning of the temporal and the spatial coordinates.  
  In this case the function $I(A,B)$ is extensive in the large $\Delta \nu$ limit since both $S(B)$ and $S(A\cup B)$ are extensive there ($S(A\cup B)$ has mean entropy per unit $\nu$ equal to $0$).      
It is remarkable that the formula (\ref{veintiuno}) shows a mutual information which is proportional to Rindler's time also for small $\Delta\nu$. The physical meaning of this fact beyond the explicit calculation in \cite{calcar} is not clear to the author. In the next section we will encounter again the two dimensional conformal mutual information (\ref{jjj}), but we will be mainly interested in the large $\Delta \nu$ region.

Now let us turn attention to Rindler space in more dimensions. 
Let us compare the naive thermal entropy of the Unruh radiation inside a given set with the corresponding mutual information between the wedge and the set. The formula (\ref{tuy}) gives the leading term for the mutual information between the wedge and a rectangular set with a large parallel face of area ${\cal A}$ and  short separating distance $R$. This gives  
\begin{equation}
I(A,B)\sim \kappa  \frac{{\cal A}}{R^{D-2}} \label{yuyu} \,.  \hspace{2cm}   
\end{equation}
On the other hand, the integration on the rectangular set of twice the large volume limit entropy density of a free massless gas gives
\begin{equation}
2 \int \frac{ds}{dV}\, dV\,\sim \hat{\kappa} \frac{{\cal A}}{R^{D-2}}\label{yoyo}\,,
\end{equation}  
where for each field degree of freedom it is 
\begin{equation}
\hat{\kappa}=\frac{ D}{(D-1)(D-2)2^{2(D-2)} \pi^{\frac{3(D-1)}{2}}\Gamma\left(\frac{D-1}{2}\right)}\int_0^\infty \frac{d\lambda \, \lambda^{D-1}}{\exp (\lambda)\pm 1}\,.
\end{equation}
The plus and minus signs are for fermions and bosons respectively. Comparing ({\ref{yuyu}) and (\ref{yoyo}) we see that the leading terms have the same functional form. However, in general $\kappa$ is greater than $\hat{\kappa}$ and the difference increases with the space-time dimension $D$. The table 1 shows some examples.

An interesting consequence of these results appear in connection to the discussions around the Bekenstein bound and the concept of unconstrained thermal radiation \cite{unconstrained}. This state of matter is supposed to maximize the entropy for a fixed temperature distribution in space. Thus the entropy in unconstrained thermal matter in the Rindler space can not be given by the integrated entropy density of a thermal gas with the local Rindler temperature. From the table 1 we see that the mutual information is always greater than twice the thermal radiation entropy, and already more than three times (twice) greater in $D=4$. According to the inequality (\ref{fofo}) this implies a corresponding inequality for the entropy, for any regularization which respects the correlations responsible for the mutual information. One could wonder if there is a deeper cause for this 'unnatural' relation between the mutual information and the large volume entropy density, and if it subsists in the presence of interactions. 

The reason for the differences with the two dimensional case are clear. In $D> 2$ we do not have any valid argument to relate the formula (\ref{yoyo}) with the physics of the problem. The large volume limit does not apply since the thermal wavelength $T^{-1}=2\pi x$ is of the order of the distance $x$ to the system boundaries. 
The argument for defining an unambiguous entropy of $B$ using the large boost time limit does not work here since the entropy we want to measure $S\sim {\cal A}/R^{D-2}$ is  smaller than the cutoff contribution due to the boundary $\sim {\cal A}/\epsilon^{D-2}$. Also, the strong subadditivity cannot be applied between time-like separated sets in dimension $D > 2$, and there is no proportionality to the Rindler time. The right way to unambiguously measure the entropy  is using the mutual information, which differs notably from the naive counting of gas entropy.

Consider now the opposite limit for $B$, that is, when its size is much smaller than the separating distance to the wedge. We take $B$ to be a sphere and keep the fields massless. For the conformal case the mutual information between two spheres must be a function of the cross ratio $\eta=\frac{(R_A+l) (l+R_B) }{l (R_A+R_B+l)}$, where $R_A$ and $R_B$ are the sphere radius and $l$ is their separating distance. Then using the eqs. (\ref{sus1}) and (\ref{sus2}) between two distant spheres and making a conformal transformation to map  the sphere $A$ to the wedge we get that $I\sim (R_B/l)^{D-2}$ for bosons and $I\sim (R_B/l)^{D-1}$ for fermions. We see the bosonic case has a different qualitative behavior than the integrated entropy density of a thermal gas, which has a functional form in this limit coinciding with the fermionic one. 
This recalls that, in contrast to the existence of an entropy density, in general we do not have extensivity for $I(A,B)$. 

However, it is the two dimensional case what we intend to extrapolate to the black hole space-time in  any dimensions.  
In the black hole analogy of the Rindler space $\Delta \nu$ must be interpreted as $\Delta t/(4M)$, where $t$
is the time as measured by the asymptotic observers and $M$ is the black hole mass.
Thus, $I(A,B)$, where $A$ is the black hole, would be
proportional to the asymptotic time, corresponding to a constant flux of Hawking radiation \cite{ch0}.

\begin{table}[t]
\centering
\begin{tabular}{|c|c|c|} \hline
$D$ & $\kappa$ & $\hat\kappa$ \\ \hline
$2$ & $1/3$ & $1/3$ \\ \hline
$3$ & $3.97 \,10^{-2}$& $2.90 \,10^{-2}$ \\  \hline
$4$ & $5.54 \,10^{-3}$ & $1.77 \,10^{-3}$\\ \hline
$5$ & $1.31 \,10^{-3}$ & $1.69 \,10^{-4}$\\ \hline
$6$ & $4.08 \,10^{-4}$ & $2.01 \,10^{-5}$\\ \hline
\end{tabular} 
\hspace{2cm}
\begin{tabular}{|c|c|c|} \hline
$D$ & $\kappa$ & $\hat\kappa$ \\ \hline
$2$ & $1/6$ & $1/6$  \\ \hline
$3$ & $3.61 \,10^{-2}$ & $2.18 \,10^{-2}$\\ \hline
$4$ & $5.38 \,10^{-3}$& $1.55 \,10^{-3}$\\ \hline
$5$ & $1.30 \,10^{-3}$& $1.58 \,10^{-4}$\\ \hline
$6$ & $4.06 \,10^{-4}$& $1.95 \,10^{-5}$\\  \hline
\end{tabular}
\caption{The coefficients $\kappa$ and $\hat{\kappa}$ for different space-time dimensions $D$. The table on the left refers to bosons and the one on the right to fermions (one field degree of freedom).}
\end{table}

\section{Evaporating black hole}
Is the semiclassical entropy of a black hole finite?. Both possible answers to this question  can be encountered in the literature, of course referring to different quantities. 

First consider the Cauchy surface $\Sigma_1$ in the figure 3(a). The space-time in this figure represents  a black hole which forms and subsequently evaporates completely. Near the horizon of a large black hole the space-time is nearly flat, and the horizon resembles a Rindler horizon. If we trace over the degrees of freedom lying inside the black hole on $\Sigma_1$, the entanglement entropy diverges with a leading term proportional to the area\footnote{One can be reassured that no global effect can change this divergent character by looking at the mutual information of two small sets on a nearly flat region on each side of the horizon. For short separating distance this mutual information increases without upper bound. The monotonicity of the mutual information shows that the conclusion is not changed for larger regions.}. 

On the other hand the arguments in \cite{argumentobek,turner} reveal a reason why the entanglement entropy of a black hole must be finite. This is because the black hole has a fixed mass, which provides a cutoff to the fluctuations responsible to the entropy. In fact the total entropy radiated by a black hole is of the order of and greater than  the Bekenstein entropy $A/(4G)$ \cite{page}, quite independently of the way the black hole was formed. If the global quantum state is pure this entropy should equal the entropy inside the black hole. In figure 3(a) this entropy corresponds to either the reduced state inside the whole black hole region (region I) or the state reduced to one of the Cauchy surfaces $\Sigma_2$ for the final region, where the black hole has evaporated completely (this is the region of the space-time space-like separated to the whole black hole). It is very remarkable to find a proper region of the space-time which has finite entanglement entropy. One could be even tempted to define a black hole as a causal region with this property. Unfortunately, the exact nature of this entanglement entropy is outside the semiclassical analysis since it depends crucially on the details of the final point of evaporation, which gives the contact surface between the black hole and the region III. It is a necessary condition for finiteness that this point has zero area for example, what is not the case if there is a remnant or in the evaporation of charged black holes in certain theories where the final state is an extremal black hole. In any case, the entropy in region III should grow as $M^2$, and even if it has a divergent part, it must be a constant independent of the initial black hole mass, and thus could in principle be isolated. 

The reason why the entropy in the whole Hawking radiation is finite while the near horizon fluctuations have infinite entanglement entropy is simply that most of these fluctuations fall back to the black hole and do not reach infinity \cite{turner}. They constitute what is called the thermal atmosphere which surrounds the horizon of the black hole.    
  
The definition of a continuous, time-dependent entropy, connecting the divergent entanglement in $\Sigma_1$ with the finite entanglement in $\Sigma_2$ may require quantum gravity effects.  These would regularize the divergent entanglement giving place to the sum of two terms for the Cauchy surface $\Sigma_1$. One is the Bekenstein Hawking black hole entropy $A/(4G)$, which is a geometric quantity depending on the horizon surface, and should come from the regularized near horizon divergent terms in the entropy. The second one would consist in the entropy contained far from the black hole in the Hawking radiation. According to the generalized second law this sum should always increase with time \cite{gsl}.  At the end of the evaporation, on $\Sigma_2$, the only remaining entropy is in the radiation, which naturally has to be greater than $A/(4G)$ for a non equilibrium process. However, there is also the controversy (information loss paradox) over the possibility that quantum gravity effects could restore the purity of the final (exterior) state \cite{infoloss}. In this case the fine-grained entropy of the total radiation would be zero (for pure initial states).     

\begin{figure}[t]
\centering
\leavevmode
\epsfysize=6cm
\epsfbox{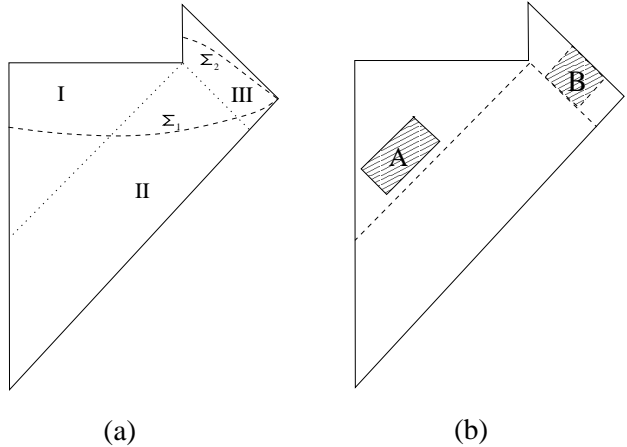}
\bigskip
\caption{(a) Two different spatial surfaces $\Sigma_1$ and $\Sigma_2$ on the space-time of a black hole which completely evaporates. (b) The set $A$ is placed inside the black hole and $B$ is in the radiation region.}
\label{f4}
\end{figure}

Therefore, at the semiclassical level there is no time dependent entropy, and at least no evident microscopic explanation to the Bekenstein-Hawking entropy (we will come back to this point in section V. See also \cite{fuuu} an references therein). Worse still, even if the total radiated entropy can be defined in the asymptotic future, in general there is no unique way to separate the entropy of the Hawking radiation in a surface like $\Sigma_1$ and it is not possible to determine where this entropy is localized. The problem resembles the one encountered for the Rindler space, but the wavelengths of the relevant modes are here much smaller than the typical distances involved. On the other hand, the gas is much more diluted than in Rindler's case, making the entropy definition also problematic. This problem of entropy localization is also present for highly diluted states and is of a general nature. We can encounter the same difficulties in the radiation of a sufficiently cold object. In order to see this, consider an object of radius $R$, and let $T$ be its surface temperature.  Take a spherical shell of radiation of outer radius $d$ which escaped during an interval of time $\Delta t< d$, much larger than the typical wavelength $T^{-1}$ of the radiation, and smaller that the cooling time.   
Assuming the radiated entropy is extensive on time (or radius) we expect the entropy in the shell to be
 $S_{\textrm{shell}} \sim T^3 R^2 \Delta t< T^3 R^2 d$, corresponding to an entropy flux in massless particles of order $T^3 R^2$. However, the regulated entanglement entropy of the shell has a contribution from the shell exterior surface which is proportional to the area in units of the cutoff squared. Choosing a distance cutoff of maximum size for the present problem, that is, of order of the radiation wavelength $T^{-1}$, the boundary contribution makes the entropy uncertain in $\delta S\sim d^2 T^2$. To extract a well defined localized entropy, it seems that we need $d\ll T \, R^2$. For a sufficiently large $d$ this inequality cannot be maintained. For example it is violated for the sun radiation for $d\gtrsim 4 \, 10^7 $ parsec and it is always violated for a black hole. One can still argue that due to the different behavior of the terms $S_{\textrm{shell}}$ and $\delta S$ with the size $ d$, it could be possible to separate them. However, even in that case, the area term gives uncertainties to the linear term in $d$. This is due to the uncertainty in the definition of $d$ which has to be of the order of our cutoff scale $\sim T^{-1}$. Thus, the naive $S_{\textrm{shell}}$ should be compared with the error in $\delta S$, of the order of $d \, T$. This gives $(RT)^2\gg 1$ in order to extract an unambiguous local entropy. This is of course not the case for the black holes which are very cold objects, having an inverse temperature bigger than their sizes.  

It is then of interest to explore the possibility that the energy flux (the energy momentum tensor) is not giving a good indicator of the entropy carried by the radiation. In the following we try to answer this question by studying the mutual information in a simple model and approximation for black hole evaporation. The mutual information gives a natural way to eliminate the area terms an divide what is nearby fluctuations from long range correlations. However, it requires specifying the other region to which the information is shared. Thus, a natural and well defined time-dependent entropy is given the mutual information between the black hole, or a part of it, and a region far away from the black hole, where the Hawking radiation is observed. The mutual information is also useful for determining where is located inside the black hole the information shared with the Hawking radiation. Thus, we consider $I(A,B)$ where a typical placing of $A$ and $B$ is as shown in fig. 3(b).

As mentioned, $I(A,B)$ is increasing with $A$ and $B$. Then it is bounded above (modulo contributions from stable remnants) by the mutual information between the whole black hole and the complete region III. If the initial state is pure this is just twice the total radiated entropy, $\lambda A/(4G)$, where $\lambda \gtrsim 1$ depends on the details of the theory. Also, the monotonicity property implies that the mutual information of the black hole and some subset of region III can only increase with the size of this later. Therefore it is not possible that correlations established during the whole evaporation period make the state in this region pure once some non zero entropy has developed at the early stages. Solutions of this type to the "information loss paradox" then would require quantum gravity. In particular, effects such as quantum fluctuations of the horizon position \cite{fluc}, or the failure of the Hilbert space in $A\cup B$ to be the tensor product of the ones in $A$ and $B$ \cite{st1}, could invalidate this reasoning. But note that ordinary QFT interactions, and in particular interactions between the outgoing modes with the falling matter, cannot.       

\subsection{Two dimensional approximation: extensivity in the radiation region}
To be concrete, in this Section we consider a spherically symmetric large black hole formed by the collapse of a thin null shell, which subsequently undergoes evaporation, as shown in the figure 4. We are interested in the relation between the black hole and the asymptotic radiation, and the issue of the final stages of evaporation will not be relevant here. 

Let us briefly describe the geometry \cite{descrip}. The spherically symmetric metric after the shell passing can be written in the quasistationary regime of evaporation with an advanced time dependent mass $m(v)\gg 1$ to a very good approximation \cite{bar} 
\begin{equation}
ds^2=\left(1-\frac{2 m(v)}{r}\right)dv^2-2dv dr-r^2 (d\theta^2+\sin^2 (\theta) d\phi^2)\label{ee}\,,
\end{equation} 
where $v$ is a null coordinate and $4\pi r^2$ measures the area of the two spheres.
As the black hole evaporates the time dependent mass obeys the energy conservation equation 
\begin{equation}
\frac{dm(v)}{dv}=-k\frac{1}{m(v)^2}\,.
\end{equation}
Here $k$ is a rate of evaporation which depends on the  field content of the theory.
Eq. (\ref{ee}) can be written in terms of null coordinates
\begin{equation}
ds^2=\left. -2\,\frac{\partial r}{\partial u}\right|_{v}\,du\,dv-r^2(u,v) (d\theta^2+\sin^2 (\theta) d\phi^2)\,,
\end{equation}
where $r(u,v)$ is given by the equations
\begin{equation}
\left. \frac{\partial r}{\partial v}\right|_{u}=\frac{1}{2} \left( 1-\frac{2 m(v)}{r}\right)
\end{equation}
and  $\left. \frac{\partial r}{\partial u}\right|_{v}=-\frac{1}{2}$ on  the asymptotically flat regions $I^+$ and $I^-$.  
 We take $v_{\textrm{shell}}=0$. The region prior to the shell transit is flat and the metric can be written as 
\begin{equation}
ds^2=dU\,dv-r^2(U,v) (d\theta^2+\sin^2 (\theta) d\phi^2)\,,
\end{equation}
with $r(U,v)=(v-U)/2$ with a different null coordinate $U$, and we have $U=v$ on $r=0$.
 The continuity of the metric along the shell $v=0$ gives the relation between null coordinates
\begin{equation}
r(u,0)=-\frac{U}{2}\,.\label{haha} 
\end{equation} 

In order to find the relation between $u$ and $U$ we find the trajectory of a light ray near the horizon. In this region we expand in 
$\delta r(u,v) = r(u,v)-2 m(v) \ll 2 m(v)$. The solution for the outgoing radial trajectory $u=$constant to leading order in $m(0)$ is exponentially increasing  
\begin{equation}
\delta r= \delta r_0 \, e^{\frac{1}{4}\int_0^v \frac{dv}{m(v)}}\,. 
\end{equation}
This exponential redshift holds roughly until the radial coordinate distance to the horizon reaches the size of the horizon radius $\delta r\sim 2 m(v)$. After that point we can use the asymptotic relation $r=(v-u)/2$.  This occurs at small $r$ compared with both the $u$ and $v$ which interest for the future asymptotically flat region,
\begin{equation}
u-v\sim 8 m(v) \ll v\,.
\end{equation} 
Thus, using (\ref{haha}), we obtain the well known exponential redshift relation between null coordinates
\begin{equation}
\log (U_h-U)= -\frac{1}{4}\int_0^u \frac{dv}{m(v)}=\frac{1}{8k} (m(u)^2-m(0)^2)\,.\label{masas}
\end{equation}
This applies in the late time region, and well before the black hole reached Planck mass size,  $e^{-\frac{m(0)^2}{8k}}\lesssim U_h-U\ll 4 \, m(0)$. 

\begin{figure}[t]
\centering
\leavevmode
\epsfysize=6.5cm
\epsfbox{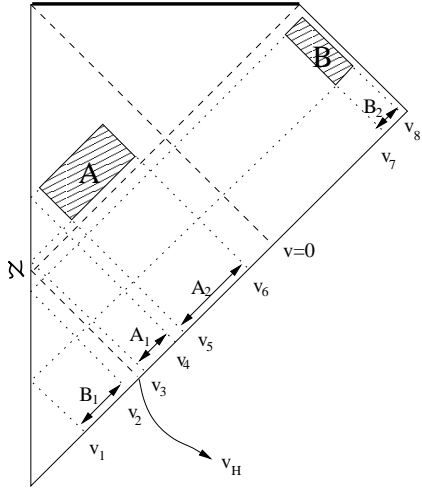}
\bigskip
\caption{Penrose diagram for the black hole formation and evaporation. In the text the mutual information $I(A,B)$ between $A$ and $B$ is calculated in a two dimensional approximation.}
\end{figure}

Now we proceed to compute the mutual information in this geometry. We will use the two dimensional approximation to black hole evaporation \cite{libro}. This consists in replacing the (massless) fields in the four geometry by a conformal field in the two dimensional geometry given by the radial-temporal part of the spherically symmetric metric. To mimic the four dimensional case a reflecting boundary condition for the two dimensional field is imposed at $r=0$.  The justification of this approximation is that the most important contribution to the Hawking radiation comes just from a few  field modes, specially the one with the lowest angular momentum. However, the conformal field in a conformally flat two dimensional metric neglects the backscattering by the geometry. Still, the effective dimensional reduction gives a good qualitative understanding of the evaporation process and of the character of information transmission as a one dimensional channel \cite{bekeinfo}. For the entanglement entropy it is not expected to give a reasonable result, since it neglects the large angular momentum modes which lead to the divergent term proportional to the area$/\epsilon^2$, with $\epsilon$ a distance cutoff (one massless field gives only a logarithmic divergence $\sim \log \epsilon$). However, we still expect it to give a reasonable qualitative approximation to the mutual information between the black hole and a distant region, since this is mainly due to the correlations between the black hole and future infinity, which are approximated by the two dimensional field. The analysis which follows is closely related to  the discussions in \cite{otroti} and \cite{vakuri}, which we hope it completes in certain aspects. In particular, our use of the mutual information justifies the claim in \cite{vakuri} that the two dimensional entanglement entropy is the relevant one for the higher dimensional case.          

The Hawking radiation is then determined by the two dimensional conformal anomaly, and the mass loss rate coefficient is given by \cite{conformal}
\begin{equation}
k=\frac{C_V}{768 \pi}\,,
\end{equation}
with $C_V$ the Virasoro central charge of the conformal theory ($C_V=1$ for a massless scalar).  

Due to the propagation along null directions in a conformal two dimensional theory the local algebras of field operators can be thought attached to regions of the null axes. In the present case, due to the reflecting boundary conditions at the origin, the two null propagating directions are mapped into one \cite{otroti}. A causal set in space-time has two past null shadows, one which goes directly to the past infinity $I^- $ and the other goes to the origin first, and then is reflected to reach $I^-$ (see figure 4). Thus to each single space-time interval it generally corresponds two disjoint segments at $I^-$ and the operator algebra corresponding to this subset of the $v$ axes. Since the state at $I^-$ is the Minkowski vacuum we can use formulas (\ref{ex}) and (\ref{chira}) (with only one chirality contributing) to evaluate the mutual information. In the example of figure 4, the quantity  $I(A,B)$ can be decomposed into four different terms of mutual information between single intervals as 
\begin{equation}
I(A,B)=I(A_1,B_1)+I(A_1,B_2)+I(A_2,B_1)+I(A_2,B_2)\,.
\end{equation}
 Under the conditions of the validity of the two dimensional approximation, which requires observing the Hawking radiation far from the black hole, the contribution from $I(A_1,B_2)$ and $I(A_2,B_2)$ can be neglected. Also, the size of $B_1$ and its $v$ coordinate distance from $v_H$  are exponentially small, which in general allows to neglect $I(A_2,B_1)$. In order to have a non negligible mutual information, $A_1$ should be also exponentially near $v_H$. Let us then consider the case where $A$ contains a piece of the horizon, $v_3=v_H$. This gives 
\begin{eqnarray}
I(A,B)=\frac{C_V}{6}\left(\log(v_H-v_1)+\log(v_4-v_2)-\log(v_H-v_2)-\log(v_4-v_1)\right) &&  \nonumber\\
\simeq 16\, \pi \left(  m(u_1)^2-m(u_2)^2) \right)&=&{\cal A}(u_1)-{\cal A}(u_2)\,,
\end{eqnarray}
up to contributions logarithmic in the black hole area ${\cal A}$. Here $u_1$ and $u_2$ are the asymptotic times of $B$,  corresponding to $U_1=v_1$ and $U_2=v_2$ through (\ref{masas}). We have assumed that the width of $A$ is not exponentially small, $v_4-v_H\gg e^{-\frac{m(0)^2}{8k}}$. 

Thus, the mutual information between the black hole and a region between two asymptotic times is proportional to the differences of black hole areas in Plank units at the corresponding times. This is four times the Bekenstein entropy. As we mentioned, one factor two comes from the definition of mutual information, which measures twice the entropy if the state in $A\cup B$ is pure. The other is due to the fact that in two dimensions the total radiated entropy in the conformal case is exactly twice the Bekenstein value (see for example \cite{vakuri}). 

Surprisingly, the mutual information turns out to be extensive on the radiation region, being distributed as dictated by the entropy density of a thermal gas with the appropriate energy flux. If this continue to hold in the full four dimensional calculation, it would reinforce the idea that it is the correct measure of localized entropy in relativity. However, the extensivity in the two dimensional problem may not give a full account of the subtle information sharing in the higher dimensional space-time. In particular, even if the correlations between $A$ and $B$ are approximately taken into account,  the field correlations in this model inside $A$ or $B$ are not realistic, and this may significantly change the results. Also, the angular distribution of information is not accessible to the two dimensional problem. 
Thus, we cannot give a definite answer to the interesting question of information extensivity here. 
 We hope to come back to this question on a future work. In any case, we remark that, contrary to the entropy, we do not know of any principle which implies spatial extensivity for the mutual information.       
 
The explicit expression for $I(A,B)$ also allows us to find the location inside the black hole of the information shared with the Hawking radiation. This is contained in any exponentially thin $\Delta U\gtrsim e^{-m(0)^2/(8k)}$, null shell, exponentially near to  the horizon, $|v_3-v_h|\lesssim e^{-m(0)^2/(8k)}$. 
No appreciable change in the mutual information is derived by including deeper regions in the black hole. Also the shell can be exponentially short in $\Delta v$ without appreciably modifying the outcome. In particular we could take $A$ to be only an exponentially small neighborhood of the point $\aleph$ where the horizon begins (see figure 4), well inside the flat space region. The mutual information of this neighborhood with all the hawking radiation would be of order ${\cal A}/G$. On the other hand eq. (\ref{fofo}) implies that the entropy in $A$ is bounded below by $I(A,B)/2$. This means that the entropy bounds get violated by a very large factor near $\aleph$ (see also the discussion in \cite{hor}).  
 
One could argue that this is no surprising since the entropy diverges semiclassically. However, the mutual information is finite and cutoff independent, and one can expect it to be stable against Plank scale physics in the macroscopic domain. The sets $A$ and $B$ can be taken macroscopic and separated by a  very large distance to each other. Thus, if the entropy bounds hold we would like to know the mechanism which makes the semiclassical calculation above not reliable. With this aim the role of backscattering is explored in the next section. Also, the interactions may impede the mode extrapolation to the $\aleph$ point, though the modes are at an extremely high frequency in this region and one should be capable to rely on the geometric optics there.   

A final commentary to this section. The mutual information forces us to think in four dimensional terms \cite{hartle}, and this brings up some common misunderstandings related to the semiclassical description of black hole evaporation.  First, the time evolution is perfectly unitary for the whole space-time, and there is no loss of information. The entropy seen for asymptotic observers is due to information sharing with the region inside the black hole. Also, sometimes it is stated that the information responsible for the entropy is destroyed at the black hole singularity. We think this is  misleading, since for the external observer the relevant information is eternally kept in the black hole interior, and even, in this dimensionally reduced calculation at least, it is retained in a small place of flat space in the region before the black hole formation. The black hole then acts as form of remnant trapped in a different time (the presence or not of a real stable remnant is then mostly irrelevant in this sense).  Third, it is sometimes argued that the entropy in the radiation evaporated from a black hole, since it depends on its whole past history, can be arbitrarily large, independently of the present size of the black hole. If the initial state is pure this entropy should then be the same as the black hole one, being inconsistent with its value being ${\cal A}/(4G)$. However, this 
again depends on what is taken as the entropy of the black hole. The mutual information shows that the correlations responsible of the Hawking radiation correspond to much earlier times, and if the black hole entropy  is related to a regularization of the surface divergent terms \cite{sorkin} (and not to all the entropy in the black hole region), there is no contradiction for it to take the Bekenstein-Hawking value.

\subsection{Localization inside the black hole: including backscattering}
In this section we include the backscattering by the Schwarzschild geometry of the four dimensional problem for free massless fields.  We focus attention on the distribution of information inside the black hole, and our primary interest is to check if a large amount of information is still contained near the neighborhood of $\aleph$ when backscattering effects are taken into account.

We use the quasistationary approximation which reduces the problem to one of a scattering in the Schwarzschild geometry of mass $M$. To introduce the notation consider first a massless scalar field. We decompose it into modes of definite angular momentum $l$ and frequency $w$ 
\begin{equation}
u=\frac{f_{wl}(r)}{r} e^{-i w t} Y_{lm}(\theta, \phi) \,.
\end{equation}
The radial function satisfies the equation 
\begin{equation}
\left(\frac{d^2}{{dr^*}^2}+w^2-V(r)\right)\,, 
\end{equation} 
with $r^*=r+2M \,\log(\frac{r-2M}{2M}) $ and the potential 
\begin{equation}
V(r)=\left(1-\frac{2M}{r}\right)\left(\frac{l(l+1)}{r^2}+\frac{2M}{r^3}\right)\,.
\end{equation}

The scattering on the potential barrier, which vanishes for $r^*\rightarrow \pm \infty$, determines the evolution of the original Unruh vacuum state. The radial function corresponding to the so called {\tiny UP} modes behaves as 
\begin{eqnarray}
f^{\textrm{{\tiny UP}}}_l(r^*, w)&\sim& e^{i w r^*} + r_l(w) e^{-i w r^*}     \hspace{1.5cm} \textrm{for} \hspace{.5cm} r^*\rightarrow -\infty \nonumber \,, \\
f^{\textrm{{\tiny UP}}}_l(r^*, w)&\sim& t_l(w)\, e^{i w r^*}\hspace{3.cm} \textrm{for} \hspace{.5cm} r^*\rightarrow \infty\,,\label{yryr}
\end{eqnarray}  
where $r_l(w)$ and $t_l(w)$ are the amplitudes for reflection and transmission through the barrier, with $|r_l(w)|^2+|t_l(w)|^2=1$. The modes relevant for the Hawking radiation for late time observers are very high frequency when propagated back to the flat region. They are entangled with the very high frequency { \tiny INT} modes which can be described by wave packets of exponentially small size entering the black hole in the region near $\aleph$. The relevant contribution to the Unruh vacuum state (which is the relevant one for black hole evaporation), in the subspace of definite energy and angular momentum, is 
\begin{equation}
\left| 0 \right>= \prod_w \sqrt{1-e^{-8 \pi M w}}\,\sum_{N=0}^\infty e^{-4 \pi M N w}  \left|N, \textrm{{\tiny UP}}\right> \otimes \left| N, \textrm{{\tiny INT}} \right>  \,,\label{cusie}
\end{equation}  
where $N$ is the occupation number and some wave packet discretization and normalization in the modes is assumed (for details see for example \cite{libro}).  

Now, according to (\ref{yryr}), the {\tiny UP} modes split into {\tiny OUT} modes, which go to infinity, and the {\tiny DOWN} modes, which fall through the horizon (at much later times than the {\tiny INT} modes). 
Accordingly, the {\tiny UP} modes write
\begin{equation}
\left|N, \textrm{\tiny UP}\right>=\sum_{p=0}^N \left( \begin{array}{c} 
N \\ P      \end{array} \right)^{\frac{1}{2}} t_l(w)^p r_l(w)^{N-p} \left|p, \textrm{\tiny OUT}\right> \otimes \left|N-p,\textrm{\tiny DOWN}\right> \,.
\end{equation}  

The relevant Hilbert space for the radiation gets then factorized as ${\cal H}={\cal H}_{\textrm{{\tiny INT}}}\otimes {\cal H}_{\textrm{{\tiny OUT}}} \otimes {\cal H}_{\textrm{{\tiny DOWN}}}$. Then, one can compute the reduced density matrices (for each mode of definite $w$ and  $l$, we suppress these indices for notational convenience)
\begin{eqnarray}
\rho_{\textrm{{\tiny INT}}}&=&\left(1-x\right)\sum_{N=0}^\infty x^{ N } \left|N,\textrm{{\tiny INT}}\right> \left< N,\textrm{{\tiny INT}}\right| \,,\\
\rho_{\textrm{{\tiny OUT}}}&=&\left(1-x\right)\sum_{N=0}^\infty \frac{(\Gamma x)^{ N }}{\left(1-(1-\Gamma) x\right)^{N+1}} \left|N,\textrm{{\tiny OUT}}\right> \left< N,\textrm{{\tiny OUT}}\right|\,,\\
\rho_{\textrm{{\tiny DOWN}}}&=&\left(1-x\right)\sum_{N=0}^\infty \frac{((1-\Gamma) x)^{ N }}{\left(1-\Gamma x\right)^{N+1}} \left|N, \textrm{{\tiny DOWN}}\right> \left< N, \textrm{{\tiny DOWN}}\right|\,,
\end{eqnarray}
with $x=e^{-8 \pi M w}$ and $\Gamma=|t_l(w)|^2$ is the transmission coefficient. 

These formulas are also valid for fields having non zero helicity, where $\Gamma$ is obtained from the adequate differential equation \cite{tt}. Also, similar expressions hold for the fermionic case. We can treat both cases, fermionic and bosonic, in a unified way by defining the functions 
\begin{equation}
s_\pm (x,\Gamma)=\pm \log(1\pm x) -\frac{1}{1\pm x} \left( \Gamma x \log (\Gamma x)\pm 
(1\pm (1-\Gamma) x) \log(1\pm (1-\Gamma) x) \right)\,,
\end{equation}
where the plus (minus) sign corresponds to fermions (bosons). Then the entropies per mode are given by
\begin{eqnarray}
S_{\textrm{{\tiny INT}}}(w,l)&=&s_\pm(x,1)\,,\\
S_{\textrm{{\tiny OUT}}}(w,l)&=&s_\pm(x,\Gamma)\,,\\
S_{\textrm{{\tiny DOWN}}}(w,l)&=&s_\pm(x,1-\Gamma)\,.
\end{eqnarray}
 
When these entropies are summed over the angular modes both, $S_{\textrm{{\tiny INT}}}$ and $S_{\textrm{{\tiny DOWN}}}$, diverge. The $S_{\textrm{{\tiny OUT}}}$ however is regulated due to the sharp fall of transmission coefficient with $l$. Thus the total radiated entropy `per unit time' can be computed \cite{page} 
\begin{equation}
\frac{dS_{\textrm{{\tiny OUT}}}}{dt}=\frac{1}{2\pi}\sum^\infty_{l=0} (2 l+1) \int_0^\infty dw \, s_\pm(x,\Gamma)\label{gggg}\,.
\end{equation}
We have raised doubts on the interpretation of this quantity as an entropy density, but  when integrated over time with a slowly varying temperature and $\Gamma$ which incorporate the change in the black hole mass during the evaporation, it must give an excellent adiabatic  approximation to the total entropy radiated.   

In consequence, to explore the distribution of information inside the black hole we cannot rely on $S_{\textrm{{\tiny INT}}}$  or $S_{\textrm{{\tiny DOWN}}}$. We can use instead the mutual informations between the asymptotic future $I^+$, with Hilbert space   ${\cal H}_{\textrm{{\tiny OUT}}}$, with the near $\aleph$ region, with Hilbert space  ${\cal H}_{\textrm{{\tiny INT}}}$, and the rest of the black hole, which we call $\hat{H}=H-\aleph $, with Hilbert space  ${\cal H}_{\textrm{{\tiny DOWN}}}$. 
The expressions of the mutual information per mode are
\begin{eqnarray}
I(\aleph,I^+)(w,l)&=&S_{\textrm{{\tiny INT}}}(w,l)+S_{\textrm{{\tiny OUT}}}(w,l)-S_{\textrm{{\tiny DOWN}}}(w,l)\,, \label{qo}\\
I(\hat{H},I^+)(w,l)&=&S_{\textrm{{\tiny DOWN}}}(w,l)+S_{\textrm{{\tiny OUT}}}(w,l)-S_{\textrm{{\tiny INT}}}(w,l)\,,\\
I(H,I^+)(w,l)&=&2 S_{\textrm{{\tiny OUT}}}(w,l)\,.\label{po}
\end{eqnarray} 

\begin{table}
\centering
\begin{tabular}{|c|c|c|c|} \hline
$h$ & $S_{\textrm{{\tiny OUT}}}/S_{\textrm{BH}}$ & $I(\aleph,I^+)/(2 S_{\textrm{BH}})$ & $I(\hat{H},I^+)/(2 S_{\textrm{BH}})$ \\ \hline
$0$ & $1.88$ & $1.49$ & $0.40$\\ \hline
$1/2$ & $1.64$ & $1.35$ & $0.29$ \\ \hline
$1$ & $1.50$ & $1.27$ & $0.23$\\ \hline
$2$ & $1.35$ & $1.19$ & $0.16$\\ \hline
\end{tabular} 
\caption{Numerical results on the mutual information between the asymptotic future $I^+$ (containing the Hawking radiation) and two regions in the interior of the black hole, the neighborhood of the point $\aleph$ where the horizon begins, and the rest of the horizon $\hat H$ (for later times). The results for the evaporation using different massless field helicity $h$ are normalized to twice the Bekenstein entropy $S_{\textrm{BH}}$. The sum of the last two columns equals the first one. This later gives the total entropy radiated and reproduces results in \cite{page}. }
\end{table}

Now, we can use the expression analogous to (\ref{gggg}) to compute the full mutual information functions
\begin{equation}
I=\int dt \,\,\frac{1}{2\pi}\sum^\infty_{l=0} (2 l+1) \int_0^\infty dw \, I(w,l)\,.
\end{equation}
This mode by mode calculation is possible because of two reasons. First, we are considering the whole asymptotic region $I^+$, so no problem with the boundaries of a finite region can arise.  Also we are considering a fixed $\aleph$ region, whose actual size, smaller than the black hole radius and bigger than exponentially small distances, is not relevant. This means that if correlations between $\aleph$ and $\hat{H}$ change their mutual informations with the distant $I^+$ region they do it by a fixed quantity, much smaller than the black hole entropy.    

The expressions (\ref{qo}-\ref{po}) can be compared with the black hole Bekenstein entropy loss due to loss of energy 
\begin{equation}
\frac{dS_{\textrm{BH}}}{dt}=4M \sum^\infty_{l=0}(2 l+1) \int_0^\infty dw \frac{w\, x\, \Gamma }{1\pm x}\,.
\end{equation}
Comparing the mode contributions we always have 
\begin{equation}
S_{\textrm{{\tiny OUT}}}\ge S_{\textrm{BH}}\label{ghgh}
\end{equation}
 as corresponds to the generalized second law \cite{page}. This is true for an arbitrary $\Gamma$ function and just expresses that the out of equilibrium radiation process  from a thermal reservoir must increase the total entropy.

Remarkably,  the total amount of mutual information with the initial aleph point is kept finite, for an exponential small set $\aleph$. This large violation of the entropy bounds is then not eased by backscattering. Moreover, we have 
\begin{equation}
I(\aleph,I^+)\ge 2 S_{\textrm{BH}}\,.\label{sese}
\end{equation} 
This is true for fields of any helicity, since 
\begin{equation}
\frac{s_\pm(x,\Gamma)+s_\pm(x,1)-s_\pm(x,1-\Gamma))}{ \Gamma x |\log (x)| /(1\pm x)}\ge 2 \label{sete}
\end{equation}
for any $x\in [0,1]$ and $\Gamma\in [0,1]$. We will discuss the thermodynamical explanation for this as well as for (\ref{ghgh}) in Section V.  

Note that since the global state is pure we have the partial extensivity 
\begin{equation}
I(H,I^+)=I(\hat{H},I^+)+I(\aleph,I^+)=2 S_{\textrm{{\tiny OUT}}}\,.
\end{equation}
This and the fact that $s_\pm(x,\Gamma)$ is increasing with $\Gamma$ lead to 
\begin{equation}
S_{\textrm{{\tiny OUT}}}\le I(\aleph,I^+)\le 2 S_{\textrm{{\tiny OUT}}}\,.
\end{equation}

Some numerical results for radiation by massless fields of different helicities in four dimensions are given in table II. The numerical methods are discussed in \cite{numer}. 
 For the two dimensional approximation $\Gamma=1$ and we have $S_{\textrm{{\tiny OUT}}}/S_{\textrm{BH}}=2$, $I(\aleph,I^+)/(2 S_{\textrm{BH}})=2$, as we found in the previous section. For a massive field with mass $m\gg (M G)^{-1}$, we have $\Gamma\simeq \theta (4 M w-l)$ and this gives almost reversible evaporation \cite{pagezur}, $S_{\textrm{{\tiny OUT}}}\simeq S_{\textrm{BH}}$, but with an exponentially suppressed flux, 
\begin{equation}
\frac{dS_{\textrm{BH}}}{dt}\simeq \frac{2}{\pi^2} m^2 M e^{-8\pi M m}  \,\label{massivo}.
\end{equation}
In this case, and in any dimension,  we have almost all the mutual information concentrated in the $\aleph$ region, $I(\aleph,I^+)\simeq 2 S_{\textrm{BH}}$. 

\subsection{Localization inside the black hole and conformal symmetry}
 
Given a pure state $\left|0\right>$ in a bipartite system ${\cal H}={\cal H}_1\otimes {\cal H}_2$ we can choose basis in order to express it in the Schmidt decomposition \cite{dieci}
\begin{equation}
\left|0\right>=\sum \lambda_i \left| \alpha_i\right>\otimes \left| \beta_i\right>\,,
\end{equation}  
 with $\sum |\lambda_i|^2 =1$, in a similar fashion as eq. (\ref{cusie}). Then there is a one parameter unitary symmetry acting on the global Hilbert space given by
 \begin{equation}
 U(s)=\rho_1^{-i s}\otimes \rho_2^{i s}\,,
 \end{equation} 
where 
\begin{eqnarray}
\rho_1=\sum |\lambda_i|^2 \left| \alpha_i\right>\otimes \left< \alpha_i \right| \,,\\
\rho_2=\sum |\lambda_i|^2\left| \beta_i\right>\otimes \left< \beta_i \right|\,,
\end{eqnarray}
are the reduced density matrices. The symmetry then leaves the state $\left|0\right>$ invariant 
\begin{equation}
U(s)\left|0\right>=\left|0\right>\,.
\end{equation}

Thus, in quantum field theory any decomposition of the space in two regions leads to a one parameter group of symmetries which leaves the vacuum invariant. This is one of the results of the Tomita-Takesaki theory \cite{tomita} (modulo details of the mathematical definitions, since strictly speaking the reduced density matrices do not exist in QFT). This unitary group is called the modular group and $s$ acts as a kind of internal time for the two regions. For any QFT, the Rindler decomposition of the Minkowski space in two half spaces leads to a modular group whose the action  is given by the boost symmetries \cite{boo} (see (\ref{seee})). In general however, the modular group does not have a simple action in terms of space-time point transformation on the field operators.

\begin{figure}[t]
\centering
\leavevmode
\epsfysize=6cm
\epsfbox{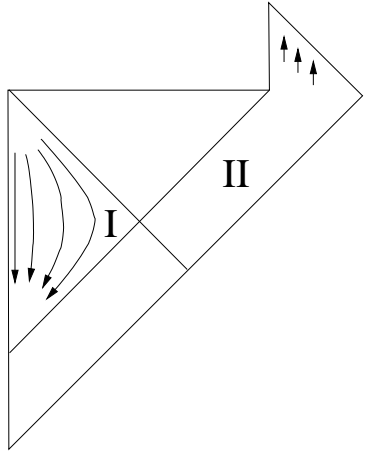}
\bigskip
\caption{Black hole formed by the collapse of a null shell. The region I below the shell and above the horizon is a diamond shaped subset (intersection of the past of a point and the future of another point) of Minkowski space. Region II contains all points space-like separated from region I. The arrows show the trajectories of the modular group. It acts geometrically in region I for conformal fields and in the late time region II if thermal equilibrium is achieved. }
\end{figure}

Now, suppose that we have a further decomposition ${\cal H}_1={\cal H}_A \otimes {\cal H}_{\bar{A}}$ and ${\cal H}_2={\cal H}_B \otimes {\cal H}_{\bar{B}}$ and we are interested in the mutual information $I(A,B)$. An element of the modular group will keep the decomposition ${\cal H}={\cal H}_1\otimes {\cal H}_2$ invariant, but will carry these new decompositions into different ones by the unitary transformation ${\cal H}_1={\cal H}_{A^\prime} \otimes {\cal H}_{\bar{A^\prime}} $ and ${\cal H}_2={\cal H}_{B^\prime}\otimes {\cal H}_{\bar{B^\prime}}$. As the global state is kept the same we have 
\begin{equation}
I(A,B)=I(A^\prime,B^\prime)\,.\label{rela}
\end{equation}

We want to use this property with the modular group corresponding to the space-time of figure 5, representing a black hole formed by the collapse of a null shell. One of the two sets in which it is decomposed the space-time corresponds to the diamond shaped set in flat space bounded by the horizon on the lower side and the incoming null shell on the upper one. As the state is the Minkowski vacuum the action of the modular group in this region is a flat space problem (it depends on $\rho_1$). In general the action of the modular transformations is not known, an exception being the case of conformal theories, in which the diamond can be transformed to the Rindler wedge. We have that the modular group in this case acts as a point transformation given 
by \cite{diamo} 
\begin{equation}
X_\pm^\prime=2M\,\left( (2M+X_\pm)-e^{2\pi s} (2M-X_\pm)\right) \left( (2M+X_\pm)+e^{2\pi s} (2M-X_\pm)\right)^{-1} \,, 
\end{equation}
where $X_+=t-r+2M$ and $X_-=r+t+2M$ are null coordinates, and $2M$ is the radius of the diamond (Schwarzschild radius). The origin of time was set such that $v=r+t=0$ for the incoming shell. Thus, we have chosen $X_+=X_-=2M$ for the upper vertex and $X_+=X_-=-2M$ for the lower diamond vertex (which we have called $\aleph$). Thus, any set $A$ included in this region will be carried to $\aleph$ to an exponentially fast velocity for large $s$
\begin{equation}
\delta X_\pm^\prime=X^\prime_\pm+ 2M \sim 4M\,\frac{2M+X_\pm}{2M-X_\pm} \, e^{-2\pi s}\,.  
\end{equation}     
This means due to (\ref{rela}) that any mutual information of a set in the diamond will also be achieved for sets as  near $\aleph$ as we want. 

The modular group would in general not act geometrically in the region opposite to the diamond (region II in figure V). However, if we can make the black hole to come to a thermal equilibrium at a temperature $T$ with the radiation in the distant future (or the radiation in thermal equilibrium alone), by imposing boundary conditions or using an asymptotically AdS space, then the modular group will act as time translations there. This is because since the reduced density matrix is thermal $\rho_2\sim e^{-H/T}$ the modular group is $U(s)=e^{-i H s/T}$. In this case, the mutual information for the exponentially small ($\sim e^{-2 \pi s}$) and exponentially near $\aleph$ set $A^\prime$ with the region $B^\prime$ time translated by an amount $\Delta t=s/T$ is the same as  $I(A,B)$ for any value of the parameter $s$ as large as we want.  Note that this relation somehow encodes the Hawking effect. If in the black hole model of Section III we keep the mass constant $m(v)\equiv M$ by providing the necessary incoming radiation, this is exactly the relation we find from the one between coordinates corresponding to (\ref{masas}) 
\begin{equation}
\log (U_h-U)= -\frac{1}{4} \frac{t}{M}=-\frac{1}{4} \frac{s}{T M}=-2 \pi s \,,
\end{equation}
 where we have used the black hole temperature $T=(8\pi M)^{-1}$.
 
\section{A finite semiclassical Bekenstein entropy and the generalized second law}

We now comment on an interesting phenomena which was pointed out in \cite{ch0} (see also \cite{aqft,kleva}) and which we interpret here as a possibility to understand the Bekenstein-Hawking entropy at the semiclassical level. This, contrary to the usual assumptions about quantum gravity, does not require a reduction of degrees of freedom, or a cutoff, at high energies, but exactly the contrary, a large number of field degrees of freedom.  
 
In order to proceed consider two sets with large parallel faces of area ${\cal A}$ at a distance which is larger than the inverse of the free field mass, $ML\gg 1$. The formula (\ref{masapan}) gives a mutual information exponentially decreasing with $L$  
\begin{equation}
I(A,B) \simeq  \frac{{\cal A}}{L^{D-2}} \frac{(ML)^{\frac{D-3}{2}}}{\pi^{\frac{D-3}{2}}2^{D+2}}e^{-2ML}\,,
\end{equation}
which is the same for each free bosonic and fermionic degree of freedom. This follows from the asymptotic form of the entropic c-functions $c(t)\sim 1/4 (n_B+n_F) t K_1(2t)$ for large $t\gg 1$, with $K_1$ the modified Bessel function \cite{ch4,ch3}.
 Suppose now that we have a series of independent massive fields and call the density of fields per unit mass $\rho(M)$. If this density is exponentially increasing at high values of the mass $\rho(M)\sim f(M) e^{M/T_H}$, it is known that the theory acquires a Hagedorn highest possible temperature $T_H$. This also implies that the mutual information 
\begin{equation}
I(A,B)\simeq {\cal A}\,\, 2^{-(D+2)}L^{-\frac{D-1}{2}}\pi^{-\frac{D-3}{2}}\int^\infty dM\,  \, M^{\frac{D-3}{2}}\,f(M) \,e^{-M\,(2 L-T_H^{-1})}
\end{equation}
diverges when the sets are at a finite distance to each other smaller than $d_H=T_H^{-1}/2$ (rather than when they are in contact).  
Remarkably,  at the critical distance the mutual information can still be finite. This depends on whether the integral of $f(M)\, M^{\frac{D-3}{2}}$ is convergent. If the energy density for the Hagedorn temperature is finite then the mutual information for critical distance is also finite. Thus, the mutual information may incorporate a natural distance cutoff $d_H$ which effectively makes $I(A,B)$  bounded for a fixed $A$ (and $B$ in the same spatial plane),  giving place to a form of entropy bound. For $B$ at distances smaller than $d_H$, strictly speaking, the mutual information cannot be defined \cite{aqft} (there is no product Hilbert space ${\cal H}_A\otimes {\cal H}_B$). 
   
We have argued above that the mutual information can represent the Hawking radiation entropy unambiguously. If there is a Hagedorn temperature, it is possible that it could incorporate also the black hole entropy. Let us think again in the Cauchy surface $\Sigma_1$ in figure 3(a). We take as $A$ the region of the black hole interior on this surface and the second set $B$ the subset of the exterior which makes $I(A,B)$ maximal. This incorporates a distance cutoff to the horizon in a way which resembles the stretched horizon or the brick wall models \cite{tho,st}. This Cauchy surface dependent mutual information $I_\Sigma$, then has a term proportional to the horizon area due to the tower of massive degrees of freedom, and also the contribution of the massless fields due to the Hawking radiation. Note that (\ref{massivo}) implies that the contribution of the massive modes to the Hawking radiation is negligible for black holes larger than the Hagedorn temperature. As a bonus, this interpretation of the black hole entropy has the advantage of not requiring any discreteness in the space-time nor in the fundamental degrees of freedom.
 
This proposal would have to be completed in two different directions. In first place there should be a connection between the Hagedorn transition and gravity in order that the coefficient of the area in the mutual information at the critical distance to be equal to $(2G)^{-1}$. We can only mention in this sense the works \cite{wo} which are also related to the connections between black holes and a Hagedorn transition. 

As a second requirement one would like to be capable to explain the generalized second law (GSL). The existent versions of proofs of the GSL at the semiclassical level obviate that the entropy diverges and cannot be well defined. The Sorkin's proof \cite{sor} rely on the fact that the time evolution outside the black hole is autonomous. This interesting proof cannot be directly translated to our candidate for the total entropy (black hole plus radiation) as half the maximum mutual information $I_\Sigma$, at least in an obvious manner, since this later also depends on the black hole interior (see \cite{argumentobek} for a related discussion).  

On the other hand, the content of the Frolov and Page proof for the quasistationary case is very similar to what is required here \cite{frol}. We recall their work. In the quasistationary regime they map the problem to one of a scattering in an eternal black hole. The two different times corresponding to $\Sigma_1$ and $\Sigma_2$ at which one should compare the entropies are taken to be the asymptotic past and future. One then makes the Cauchy surface partition $\Sigma_1=H^- \cup I^-$ and $\Sigma_2=H^+\cup I^+$, with $H^+$ and $H^-$ the future and past horizons respectively. Then, they argue that the initial state should be taken as uncorrelated, \begin{equation}
\rho_{\Sigma_1}=\rho_{H^-}\otimes \rho_{I^-}\,,\label{pr}
\end{equation}
 a key assumption which usually makes the second law work.  For the final state we have by the unitary evolution 
$ S(\Sigma_2)=S(\Sigma_1)=S(H^-)+S(I^-)$,   and by subaditivity of the entropy
\begin{equation}
S(H^+)+S(I^+)\ge S(\Sigma_2)\label{se}\,.
\end{equation}
Then we have
\begin{equation}
S(H^+)+S(I^+)\ge S(H^-)+S(I^-)\,.\label{ter}
\end{equation}    
Now, one introduces elements from the quasistationary nature of the evaporation. The emitted state $\rho_{H^- }$ is taken thermal at the black hole temperature $T$, and the Hilbert space on $H^+$ and $H^-$ are assumed to be equivalent, mapped to each other by the CPT operator $\Theta$ of the eternal black hole geometry, which commutes with the Hamiltonian. This implies that we can use the positivity of $S(\rho_{H^-}|\Theta \rho_{H^+} \Theta^\dagger )$ to show that the free energy is a minimum for the thermal state $\rho_{H^-}$
\begin{equation}
\frac{E^-}{T}-S(H^-)\le \frac{E^+}{T}-S(H^+)  \,,\label{cuars}
\end{equation}
where $E^+$ and $E^-$ are the incoming and outgoing energies respectively.  
 Assuming the usual thermodynamical equilibrium relations hold between entropy, energy and temperature of the black hole and the quasistationary approximation gives
 \begin{equation}
\Delta S_{BH}=\frac{E^+ - E^-}{T}\,.
 \end{equation}
 Then, the combination of the inequalities (\ref{ter}) and (\ref{cuars}) give the GSL,
 \begin{equation}
 S_{I^+}-S_{I^-}+\Delta S_{BH}\ge 0\,.\label{ggssll}
 \end{equation}
 Note that this demonstration makes no particular use of the specific expression of the black hole entropy, and only rely on its character of thermal reservoir. This fact underlies the general validity of the inequality (\ref{ghgh}) \cite{frol}. 
 
 In terms of the mutual information the equation (\ref{pr}) reads $I(H^-,I^-)=0$, and the eq. (\ref{se}) is $I(H^+,I^+)\ge 0$. Then (\ref{ter}) means
\begin{equation}
I(H^+,I^+)\ge I(H^-,I^-), 
\end{equation}  
which can be taken as a quasistationary idealization for to the proposal here $I_{\Sigma_2}\ge I_{\Sigma_1}$. This equation just expresses the development of correlations between the system and the hidden sector (the black hole here), which is the mechanism of the second law.

The results of the previous section suggest a different formulation of the second law. This requires interpreting $I(A,B)$ as the sum of the black hole plus exterior entropy, but where now $A$ is kept fixed as a neighborhood of the $\aleph$ point, and $B$ is again the subset of a Cauchy surface $\Sigma$ exterior to the black hole and which maximizes the mutual information. This formulation has the advantage of making explicit a difference between past and future, which is present in the second law. Presumably, the contribution from the massive modes would also be proportional to the black hole area on $\Sigma$ in this case. In the quasistationary scheme of Frolov and Page we have to take as $\aleph$ the hidden sector to which the thermal state  $\rho_{H^-}$ is entangled, or, more generally, a sector which allows the purification of the total state, that is, such that $\rho_{I^-\cup H^-\cup \aleph}$ is pure. Thus we have  $S(\aleph \cup I^-)=S(H^-)$ and, assuming that the time evolution does not mixes the $\aleph $ sector with the rest,  $S(\aleph \cup I^+)=S(H^+)$. Using this and the inequalities (\ref{cuars}) and (\ref{ggssll}) the  GSL in this new formulation follows
\begin{equation}
I(\aleph,I^+)-I(\aleph,I^-)+2\,\Delta S_{BH}\ge 0\,,\label{verssion}
\end{equation}    
where, as above $\Delta S_{BH}=\Delta E/T$ takes its quasistationary value (in the full version it would be replaced by the contribution of the massive modes to $I(\aleph,I^+)$ and $I(\aleph,I^-)$). Again, this also holds in a more general context, since we only have taken into account the black hole as a thermal reservoir which is initially uncorrelated to the exterior system. This is behind the general validity of (\ref{sese}) and (\ref{sete}).

\section{Final comments}
It is generally believed that in the physical applications of the concept of entropy it must not be taken as an absolute quantity, but it has to be understood with reference to some form of coarse graining. In practice however, this later is rarely spelled out. It is therefore satisfying that the combination of relativity and quantum mechanics might impose one such specific form of coarse graining, only allowing to talk of the entropy enclosed in a region while in relation with another one, in the form of shared information. 

Indeed, the mutual information seems to be the right tool to deal with the vacuum fluctuations which plague of divergences the localized entropy in QFT. This is specially relevant for dilute entropy scenarios such as the Hawking radiation escaping an evaporating black hole. The calculations done so far in this paper shed only a partial light on this issue and we have certainly asked more questions than obtained answers.

One important point to be further explored is the extensivity of $I(A,B)$ in the radiation region, which is suggested by the distribution of energy, but does not seem to follow from a simple derivation. 
In this sense, it would also be relevant to find out the conditions implying the extensivity of the mutual information, $I(A,B,C)=0$, in a general quantum system. In particular, we do not have a simple understanding of the exact extensivity in conformal theories. One may wonder if there is a deeper connection between extensivity of the mutual information in the radiation region in the full higher dimensional problem and two dimensional conformal theory. 

If the mutual information can replace the notion of entropy, the extensivity may also  be relevant in a different direction. Suppose an entropic object falls to the black hole. In order to include its entropy in the scheme one should express it as mutual information with a hidden sector $A^\prime$ (which can also be another black hole or a cosmological horizon). Then, it is natural to expect that the mutual information of the exterior region $B$ with the black hole $A$ plus the hidden sector, $I(A\cup A^\prime,B)$, separates as a sum $I(A,B)+I(A^\prime,B)$. One can assume that $A$ and $A^\prime$ are uncorrelated between themselves, but the reason why the correlations of these degrees of freedom with a localized $B$ would have additive information is not clear.   

Another subject of interest concerns the stability properties of $I(A,B)$ in QFT against changes in the high energy sector of the theory. This is relevant regarding the information loss paradox. If information is retrieved in the Hawking radiation one must understand through which mechanism the indications of this paper fail: monotonicity in the radiation region and a large information gathering in the neighborhood of $\aleph$. 

We think it is also very important to understand the black hole entropy at the semiclassical level. This is especially relevant   since there has not been a systematic investigation up to present as to whether quantum gravity is a physically decidable theory. For example, according to one school, effects of quantum gravity lead to the information recovery in the Hawking radiation after complete black hole evaporation. However, even if this may hold in a particular model, it would not be possible to check it experimentally. This is not due to practical reasons, but because of fundamental ones, since gravitons are inevitably emitted, and there is not enough matter to fully measure them  in an asymptotically flat (or de Sitter) space \cite{smol}.  

\section{Acknowledgments}
I thank J. Russo and M. Huerta for discussions.

\end{document}